\documentclass[12pt]{article}
\usepackage{epsfig}
\usepackage{amsfonts}
\usepackage{latexsym}
\usepackage{amsmath,amssymb}
\usepackage{mathrsfs}
\usepackage{hyperref}
\usepackage{setspace}
\usepackage{color}
\usepackage{bm}
\numberwithin{equation}{section}

\begin{document}

\begin{titlepage}
\unitlength = 1mm
%\today 

\vskip 1cm
\begin{center}

 {\Large {\textsc{\textbf{Cosmological implications of quantum entanglement \\in the multiverse}}}}

\vspace{1.8cm}
Sugumi Kanno

\vspace{1cm}

{\it $^*$ Department of Theoretical Physics and History of Science,
University of the Basque Country UPV/EHU,
48080 Bilbao, Spain}

\vspace{0.2cm}

{\it $^\flat$ IKERBASQUE, Basque Foundation for Science, 
Maria Diaz de Haro 3,
48013, Bilbao, Spain}

\vspace{0.2cm}

\vskip 1.5cm

\begin{abstract}
\baselineskip=6mm
We explore the cosmological implications of quantum entanglement between two causally disconnected universes in the multiverse. We first consider two causally separated de Sitter spaces with a state which is initially entangled. We derive the reduced density matrix of our universe and compute the spectrum of vacuum fluctuations. We then consider the same system with an initially non-entangled state. We find that scale dependent modulations 
may enter the spectrum for the case of initially non-entangled state due to quantum interference. This gives rise to the possibility that the existence of causally disconnected universes may be experimentally
tested by analyzing correlators in detail.
\end{abstract}

\vspace{1.0cm}

\end{center}
\end{titlepage}

\pagestyle{plain}
\setcounter{page}{1}
\newcounter{bean}
\baselineskip18pt

\setcounter{tocdepth}{2}

\tableofcontents

\section{Introduction}
Inflationary cosmology and string landscape suggest that our universe may not be the only universe but is part of a vast complex universes that we call the multiverse~\cite{Sato:1981gv, Vilenkin:1983xq, Linde:1986fc, Linde:1986fd, Bousso:2000xa, Susskind:2003kw}. 
This multiverse idea has been a source of debate for a last few decades and criticized as a philosophical proposal that cannot be tested. However, the idea may be validated by using quantum entanglement between two causally separated universes in the multiverse. In the multiverse, an infinite number of local universes with enormous diversity of environments can be produced once inflation happens. In such a situation, our universe may have been nucleated as an entangled pair of universes coherently under certain causal mechanism, and then may have been separated off by the exponential expansion of space between us. Our parter universe may exist beyond the Hubble horizon afterwards, and the quantum fluctuations of our universe may be entangled with those of the unobservable universe out there. Some effects of the entanglement with unobservable universe may appear in the CMB of our universe. 

Quantum entanglement is a counterintuitive phenomenon predicted by quantum mechanics. It has fascinated many physicists since 1935 when Einstein-Podolsky-Rosen (EPR) pointed out that performing a local measurement may affect the outcome of local measurements instantaneously beyond the light cone~\cite{Einstein:1935rr}. In 1981, Aspect et al performed a convincing test that the quantum entanglement is a fundamental aspect of quantum mechanics by measuring correlations of linear polarizations of pairs of photons~\cite{Aspect:1981zz, Aspect:1982fx}. After the experiment, more attention has been paid to how to make use of quantum entanglement of EPR pairs in quantum cryptography and quantum teleportation (see \cite{Horodecki:2009zz} and references therein).

Recently, in the course of progress in string theory, Maldacena and Pimentel developed an explicit method to calculate the entanglement entropy in a quantum field theory in the Bunch-Davies vacuum in de Sitter space~\cite{Maldacena:2012xp}. They showed that quantum entanglement can exist between two causally disconnected regions in de Sitter space. Then~\cite{Kanno:2014ifa} examined the spectrum of vacuum fluctuations by using the reduced density matrix derived by them and found that the quantum entanglement affects the shape of the spectrum on large scales comparable to or greater than the curvature radius.

If the quantum state of our universe was entangled initially with an unobservable universe out there, it would be reasonable to discuss its effects on the spectrum. 
In fact, the research of~\cite{Kanno:2014bma} showed that the entanglement of an initial state between two causally disconnected de Sitter spaces may remain on small scales.
Since we have no access to the causally disconnected universes, we need to trace out the inaccessible universes and thus lose information about them. Then the quantum mechanical system consisting of our universe and the rest of unobservable universes becomes a mixed state, if the quantum state is entangled initially. On the other hand, if the initial quantum state is not entangled, the quantum mechanical system remains a pure state even if we trace out the unobservable universes. As a consequence, quantum interference may play an important role. This difference of the initial state reflects on the power spectrum.

For simplicity, in this paper, we consider two universes in the inflationary multiverse. 
Assuming that one of them is our universe, we demonstrate how the difference between
entangled and non-entangled initial states reflects on the spectrum of 
vacuum fluctuations.

The paper is organized as follows. In Section~\ref{s2}, we consider initially entangled two de Sitter spaces and derive the reduced density matrix of our universe. As a simple example, we consider an entangled state between the vacuum and the one-particle state. We calculate the spectrum of vacuum fluctuations by using the reduced density matrix. In Section~\ref{s3}, we consider the same system with an initially non-entangled state. We compute the spectrum and then compare it with that of the initially entangled state. We see that 
quantum interference distinguishes the spectrum of the entangled state from the one with the non-entangled state. In Section~\ref{s4}, we present another example with an initially entangled state between the vacuum and the two-particle state. We compare it with the
corresponding non-entangled state and demonstrate that the quantum interference contributes to the spectrum directly for the non-entangled state. 
Our results are summarized and discussed in Section~\ref{s5}. 
Some technical details are presented in the Appendix~\ref{appA}.

\section{Spectrum for initially entangled state}
\label{s2}
\begin{figure}[t]
\vspace{-2cm}
\includegraphics[height=9cm]{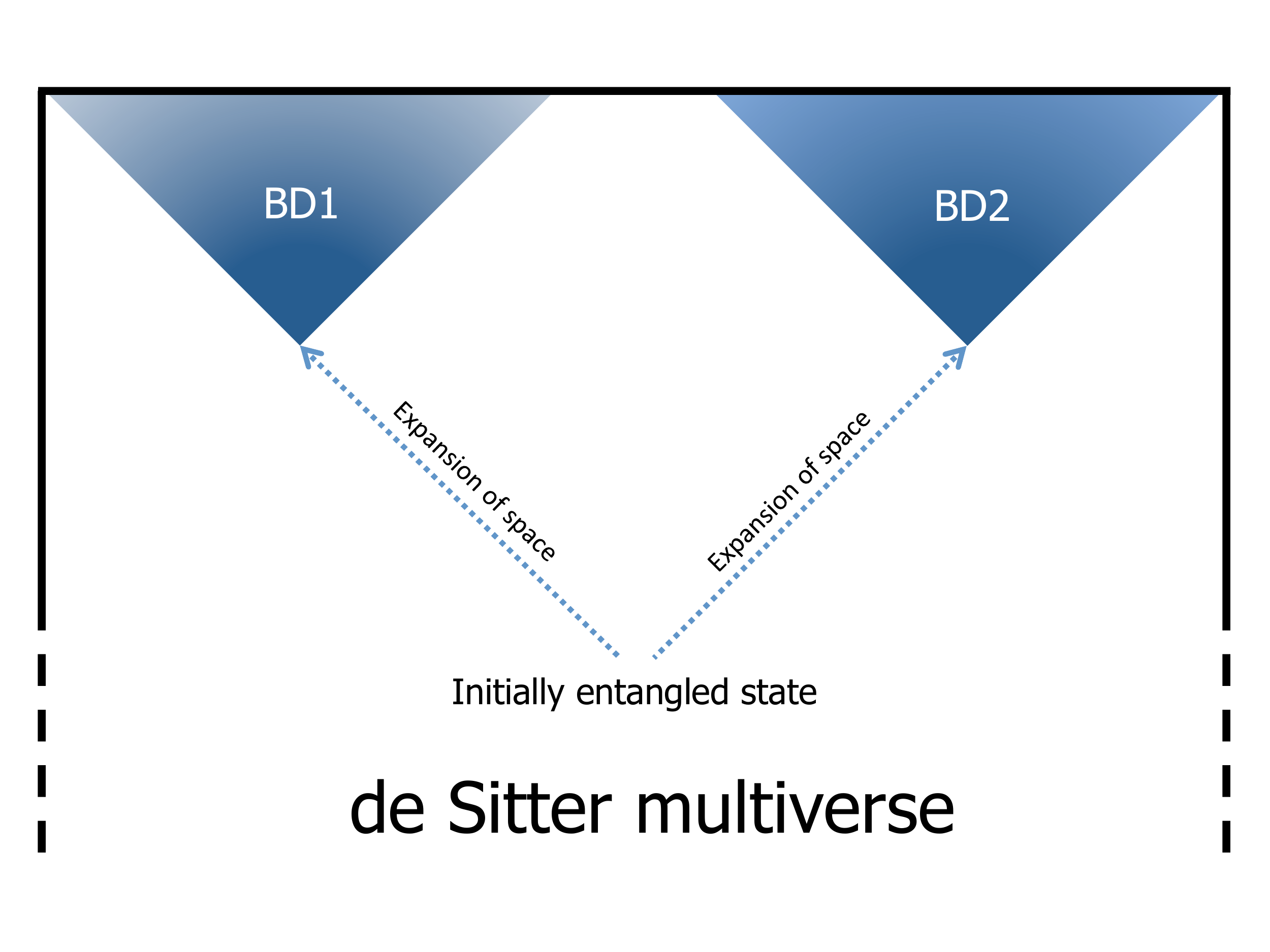}\centering
\vspace{0cm}
\caption{Causal diagram of the inflationary multiverse.}
\label{fig1}
\end{figure}

In this section, we discuss quantum entanglement between two causally disconnected de Sitter spaces (BD1 and BD2) depicted in Figure~\ref{fig1}. A quantum mechanical system consists of subsystems BD1 and BD2. The Hilbert space is a direct product ${\cal H}={\cal H}_{\rm BD1}\otimes{\cal H}_{\rm BD2}$. We suppose that our universe is, say, BD2 and we have no access to BD1.

\subsection{Mode functions in de Sitter space}
\label{s2.1}
In this subsection, we consider mode functions in our universe (BD2). For simplicity, we consider a massless minimally coupled scalar filed in de Sitter space. We expand the scalar field as
\begin{eqnarray}
\phi(x)=\sum_{\bm k}\phi_{\bm k}(\eta)e^{i{\bm k}\cdot{\bm x}}\,,
\end{eqnarray}
where $\eta$ is the conformal time and
\begin{eqnarray}
\phi_{\bm k}=a_{\bm k}u_k(\eta)+a_{-{\bm k}}^\dag u_k^*(\eta)\,,\qquad
\left[a_{\bm k}, a_{\bm p}^\dag\right]=\delta_{\bm k, \bm p}\,.
\end{eqnarray}
A vacuum state for BD2 is a state annihilated by $a_{\bm k}$
\begin{eqnarray}
a_{\bm k}|0\rangle_{\rm BD2}=0\,.
\end{eqnarray}
The mode function $u_{k}$ satisfies
\begin{eqnarray}
u_k^{\prime\prime}+2{\cal H}u_k^\prime+k^2u_k=0\,,\qquad
u_ku_k^*{}'-u_k^*u_k'=\frac{i}{a^2}=i(H\eta)^2\,.
\end{eqnarray}
Here a prime denotes derivative with respect to the conformal time $\eta$ and ${\cal H}=a^\prime/a$ where $a$ is the scale factor.  The action for the mode $\bm k$ is, 
\begin{eqnarray}
S_{\bm k}=\frac{1}{2}\int d\eta\,a^2(|\phi_{\bm k}'|^2-k^2|\phi_{\bm k}|^2)\,.
\end{eqnarray}
The standard positive frequency function at $\eta\to-\infty$ is
\begin{eqnarray}
u_k=\frac{H}{\sqrt{2k^3}}(1+ik\eta)e^{-ik\eta}\,.
\label{mf}
\end{eqnarray}
The following discussion, however, is independent of the choice of $u_k$.

\subsection{Initially entangled state}
\label{s2.2}

We consider an initial state where the vacuum and one-particle states are entangled,
as a simple example. Such an entangled state may arise if a pair of the universes (BD1 and BD2) are nucleated coherently under certain causal mechanism.
We focus on a single mode with indices ${\bm k}$ of the scalar field as an initial state:
\begin{eqnarray}
|\Psi\rangle={\cal C}_{\bm k}\,\Bigl(\,|0_{\bm k}\rangle_{\rm BD1}|0_{\bm k}\rangle_{\rm BD2}+{\cal E}_{\bm k}\,|1_{\bm k}\rangle_{\rm BD1}|1_{\bm k}\rangle_{\rm BD2}\,\Bigr)\,,
\label{ini1}
\end{eqnarray}
where the states $|0_{\bm k}\rangle_{\rm BD1}$ and $|1_{\bm k}\rangle_{\rm BD1}$ 
are the vacuum and the one-particle states of the mode ${\bm k}$ in BD1 and 
similarly for BD2. This initial state corresponds to truncation of an 
entangled state constructed by making use of the Bogoliubov transformation 
as explained in Appendix \ref{appA}.
${\cal C}_{\bm k}^2$ is the probability to observe the state $|0_{\bm k}\rangle_{\rm BD1}|0_{\bm k}\rangle_{\rm BD2}$. ${\cal E}_{\bm k}$ is defined to conserve the probability
\begin{eqnarray}
{\cal E}_{\bm k}=e^{i\theta}\sqrt{\frac{1}{|{\cal C}_{\bm k}|^2}-1}\,,\qquad 0<{\cal C}_{\bm k}<1\,,
\end{eqnarray}
where $e^{i\theta}$ is a phase factor.
When ${\cal C}_{\bm k}=1/\sqrt{2}$, the state becomes maximally entangled state. 
The Bunch-Davies vacuum is recovered both for BD1 and BD2 when 
${\cal C}_{\bm k}=1$ and then ${\cal E}_{\bm k}=0$. In general,
 both ${\cal C}_{\bm k}$ and ${\cal E}_{\bm k}$ depend on the wavenumber ${\bm k}$.
 For simplicity, we omit the indices ${\bm k}$ unless there may be any confusion below.

\subsection{Reduced density matrix}
\label{s2.3}
Since we have no access to the causally separated BD1, we have to trace out the degrees of freedom of BD1. Thus we effectively lose information about the inaccessible BD1. 
We then derive the reduced density matrix of BD2,
\begin{eqnarray}
\rho&=&{\rm Tr}_{\rm BD1}|\Psi\rangle\langle\Psi|
=\sum_{m=0}^\infty{}_{\rm BD1}\langle m|\Psi\rangle\langle\Psi|m\rangle_{\rm BD1}
\nonumber\\
&=&|{\cal C}|^2\,\Bigl(\,|0\rangle_{\rm BD2}\,{}_{\rm BD2}\langle 0|
+|{\cal E}|^2\,|1\rangle_{\rm BD2}\,{}_{\rm BD2}\langle 1|\,\Bigr)\,,
\label{dm}
\end{eqnarray}
where $|{\cal C}|^2|{\cal E}|^2=1-|{\cal C}|^2$. We note that our quantum
 mechanical system becomes a mixed state by tracing out the inaccessible BD1 
if the state is entangled.

\subsection{Spectrum of quantum fluctuations}
\label{s2.4}
Focusing on a single mode with indices ${\bm k}$, the the spectrum of vacuum fluctuations is calculated as
\begin{eqnarray}
{\cal P}(k, \eta)&=&\frac{k^3}{2\pi^2}{\rm Tr}\,\rho\,\phi_{\bm k}\,\phi_{\bm k}^\dag\,,\nonumber\\
&=&\frac{k^3}{2\pi^2}\left[\,|{\cal C}|^2{}_{\rm BD2}\langle 0|\phi_{\bm k}\,\phi_{\bm k}^\dag|0\rangle_{\rm BD2}
+\left(1-|{\cal C}|^2\right){}_{\rm BD2}\langle 1|\phi_{\bm k}\,\phi_{\bm k}^\dag|1\rangle_{\rm BD2}
\,\right]\nonumber\\
&=&\frac{k^3}{2\pi^2}\Bigl[\,|{\cal C}|^2|u_k|^2+3\left(1-|{\cal C}|^2\right)|u_k|^2
\,\Bigr]\,.
\end{eqnarray}
If we use Eq.~(\ref{mf}) for the mode functions, the spectrum after horizon exit ($\eta\rightarrow 0$) is evaluated as
\begin{eqnarray}
{\cal P}(k, \eta)&\xrightarrow{\eta\rightarrow 0}&\left(3-2|{\cal C}|^2\right)\left(\frac{H}{2\pi}\right)^2\,.
\label{sp1}
\end{eqnarray}
We see that the spectrum of the Bunch-Davies vacuum is recovered when ${\cal C}=1$. The amplitude of spectrum is enhanced by factor $2$ when the initial state is maximally entangled (${\cal C}=1/\sqrt{2}$).

\section{Spectrum for non-entangled state}
\label{s3}
As we saw in the previous section, the quantum mechanical system becomes a 
mixed state by tracing out the inaccessible BD1 if the state is entangled. 
The reduced density matrix turned out to consist of only diagonal elements. 
In this section, we consider an initially non-entangled state and will see that
the off-diagonal elements, which describe quantum interference,
appear in the reduced density matrix, which may produce a modulation 
in the spectrum.

\subsection{Initially non-entangled state}
\label{s3.1}
Since the one-particle states tend to enhance the spectrum compared with that of 
the Bunch-Davies vacuum, the difference between the spectra for
the entangled state and the vacuum is not solely due to the entanglement.
Therefore, in order to clearly see the effect of entanglement, we consider
a non-entangled state that includes one-particle states. 
Specifically we consider a state given by a direct product form
\begin{eqnarray}
|\Psi_0\rangle={\cal C}\,\bigl(\,|0\rangle+{\cal E}\,|1\rangle\,\bigr)_{\rm BD1}\otimes
{\cal C}\,\bigl(\,|0\rangle+{\cal E}\,|1\rangle
\,\bigr)_{\rm BD2}\,.
\label{ini2}
\end{eqnarray}
This state gives the same probability for the vacuum and one-particle states
as that for the entangled state~(\ref{ini1}), if one focuses on either of the 
universe.
This case contains extra two states $|1\rangle_{\rm BD1}|0\rangle_{\rm BD2}$ 
and $|0\rangle_{\rm BD1}|1\rangle_{\rm BD2}$, compared with Eq.~(\ref{ini1}). 

\subsection{Spectrum of quantum fluctuations}
\label{s3.2}
After tracing out the state of BD1 as we did in Eq.~(\ref{dm}), those extra two states contribute to the spectrum of the form
\begin{eqnarray}
|{\cal C}|^2\,{\cal E}^*\,{}_{\rm BD2}\langle 1|\phi_{\bm k}\,\phi_{\bm k}^\dag|0\rangle_{\rm BD2}\,,\qquad
|{\cal C}|^2\,{\cal E}\,{}_{\rm BD2}\langle 0|\phi_{\bm k}\,\phi_{\bm k}^\dag|1\rangle_{\rm BD2}\,.
\label{qi01}
\end{eqnarray}
Since the number of operators in the state does not match with that of operators $\phi_{\bm k}\,\phi_{\bm k}^\dag$, those off-diagonal elements vanish. Then, the spectrum for the non-entangled state appears to be the same as Eq.~(\ref{sp1}). However, in this case, those off-diagonal elements survive as ${}_{\rm BD2}\langle 1|\phi_{\bm k}|0\rangle_{\rm BD2}=u_{k}^*$ and ${}_{\rm BD2}\langle 0|\phi_{\bm k}|1\rangle_{\rm BD2}=u_{k}$, that is, the one point function gives a nonzero value $\langle\Psi_0|\,\phi_{\bm k}\,|\Psi_0\rangle=|{\cal C}|^2\left({\cal E}u_k+{\cal E}^*u_k^*\right)$. Thus, we need to calculate the spectrum with the variance 
$\langle(\,\phi_{\bm k}-\langle\phi_{\bm k}\rangle)(\,\phi_{\bm k}^\dag-\langle\phi_{\bm k}^\dag\rangle)\rangle$ for the non-entangled state. The square of the mean value is given by
\begin{eqnarray}
\langle\phi_{\bm k}\rangle\langle\phi_{\bm k}^\dag\rangle&=&\langle\Psi_0|\phi_{\bm k}|\Psi_0\rangle
\langle\Psi_0|\phi_{\bm k}^\dag|\Psi_0\rangle\nonumber\\
&=&|{\cal C}|^4
\left(2|{\cal E}|^2|u_{k}|^2+{\cal E}^{2}u_{k}^2
+{\cal E}^{*2}u_{k}^{*2}\right)\,.
\end{eqnarray}
The resultant spectrum becomes different from the one of entangled state Eq.~(\ref{sp1}), which is expressed as
\begin{eqnarray}
{\cal P}_0(k, \eta)
&=&\frac{k^3}{2\pi^2}\Bigl[\,\Bigl\{3-2|{\cal C}|^2
\left(1+|{\cal C}|^2|{\cal E}|^2\right)\Bigr\}|u_{k}|^2
-|{\cal C}|^4\left({\cal E}^{2}\,u_{k}^2+{\cal E}^{*2}\,u_{k}^{*2}\right)
\,\Bigr]
\nonumber\\
&\xrightarrow{\eta\rightarrow 0}&\Bigl[\,3-2|{\cal C}|^2
\left(1+|{\cal C}|^2|{\cal E}|^2\right)
-|{\cal C}|^4\left({\cal E}^{2}+{\cal E}^{*2}\right)\,\Bigr]\left(\frac{H}{2\pi}\right)^2\,.
\label{sp11}
\end{eqnarray}
We find the Bunch-Davies vacuum is recovered for ${\cal C}=1$.

The off-diagonal elements represent quantum interference. Thus, in this case, 
the quantum interference contributes to the spectrum in the form of variance,
 which turns out to distinguish the spectrum of entangled state from the one
 of the non-entangled state.

If we observe, further, the bispectrum or the higher order correlations such as $\phi_{\bm k}^n$ ($n=3$ for the bispectrum), the effect of quantum interference would contribute directly not only in the form of the variance and distinguish the spectrum between entangled and non-entangled states. We will see such a direct contribution due to quantum interference next by considering the two-particle states as a simple demonstration.

\section{Another example with two-particle states}
\label{s4}
If a pair of universes, each with one-particle states, may be nucleated, it would be 
possible to nucleate a pair of universes, one in vacuum and
the other with two-particle states. In this section, we consider an initially entangled state between the vacuum and the two-particle state, which  represents that the two-particle state is measured in either BD1 or BD2 with the probability $1/2$ (${\cal C}=1/\sqrt{2}$, ${\cal E}=e^{i\theta}$).
\begin{eqnarray}
|\Psi\rangle=\frac{1}{\sqrt{2}}\,\Bigl(\,|0\rangle_{\rm BD1}|2\rangle_{\rm BD2}+e^{i\theta}\,|2\rangle_{\rm BD1}|0\rangle_{\rm BD2}\,\Bigr)\,.
\label{ini3}
\end{eqnarray}
The spectrum after tracing out the BD1 is calculated as
\begin{eqnarray}
{\cal P}(k, \eta)&=&\frac{k^3}{2\pi^2}{\rm Tr}\,\rho\,\phi_{\bm k}\,\phi_{\bm k}^\dag\,,\nonumber\\
&=&\frac{k^3}{2\pi^2}\,\frac{1}{2}\left(\,{}_{\rm BD2}\langle 0|\phi_{\bm k}\,\phi_{\bm k}^\dag|0\rangle_{\rm BD2}
+{}_{\rm BD2}\langle 2|\phi_{\bm k}\,\phi_{\bm k}^\dag|2\rangle_{\rm BD2}\,\right)\nonumber\\
&=&\frac{3k^3}{2\pi^2}\,|u_{k}|^2\,,
\end{eqnarray}
where we used
\begin{eqnarray}
 {}_{\rm BD2}\langle 0|\phi_{\bm k}\,\phi_{\bm k}^\dag|0\rangle_{\rm BD2}=|u_{k}|^2\,,\qquad
{}_{\rm BD2}\langle 2|\phi_{\bm k}\,\phi_{\bm k}^\dag|2\rangle_{\rm BD2}=5\,|u_{k}|^2.
\end{eqnarray}
The corresponding non-entangled state with a direct product form is expressed as
\begin{eqnarray}
|\Psi_0\rangle=\frac{1}{\sqrt{2}}\,\Bigl(\,|0\rangle+e^{i\theta}\,|2\rangle\,\Bigr)_{\rm BD1}\otimes
\frac{1}{\sqrt{2}}\,\Bigl(\,|2\rangle+e^{i\theta}\,|0\rangle
\,\Bigr)_{\rm BD2}\,.
\label{ini4}
\end{eqnarray}
We trace out the BD1 similarly. Then, in this case, the extra two terms in the non-entangled state are found to produce the off-diagonal elements
\begin{eqnarray}
{}_{\rm BD2}\langle 2|\,\phi_{\bm k}\,\phi_{\bm k}^\dag\,|0\rangle_{\rm BD2}
=\sqrt{2}\,u_{k}^{*2}\,,\qquad
{}_{\rm BD2}\langle 0|\,\phi_{\bm k}\,\phi_{\bm k}^\dag\,|2\rangle_{\rm BD2}
=\sqrt{2}\,u_{\bm k}^{2}\,.
\label{qi02}
\end{eqnarray}
Thus we find that the quantum interference enters into the spectrum of non-entangled state. The spectrum is then computed as
\begin{eqnarray}
{\cal P}_0(k, \eta)
&=&\frac{k^3}{2\pi^2}\,\frac{1}{2}\left({}_{\rm BD2}\langle 0|\phi_{\bm k}\,\phi_{\bm k}^\dag|0\rangle_{\rm BD2}
+{}_{\rm BD2}\langle 2|\phi_{\bm k}\,\phi_{\bm k}^\dag|2\rangle_{\rm BD2}
\right.\nonumber\\
&&\left.\hspace{1.5cm}
+e^{-i\theta}\,{}_{\rm BD2}\langle 2|\,\phi_{\bm k}\,\phi_{\bm k}^\dag\,|0\rangle_{\rm BD2}
+e^{i\theta}\,{}_{\rm BD2}\langle 0|\,\phi_{\bm k}\,\phi_{\bm k}^\dag\,|2\rangle_{\rm BD2}
\,\right)\nonumber\\
&=&\frac{k^3}{2\pi^2}\left[\,3\,|u_{k}|^2
+\frac{1}{\sqrt{2}}\left(e^{-i\theta}\,u_{k}^{*2}
+e^{i\theta}\,u_{k}^{2}\,\right)\right]\,.
\label{sp2}
\end{eqnarray}
Since $u_{k}^2=u_{k}^{*2}=|u_{k}|^2=H^2/(2k^3)$ after horizon exit ($\eta\rightarrow 0$), the spectrum is evaluated as
\begin{eqnarray}
{\cal P}_0(k, \eta)\xrightarrow{\eta\rightarrow 0}
\left[\,3
+\sqrt{2}\cos\theta\,\right]\left(\frac{H}{2\pi}\right)^2\,.
\end{eqnarray}
An oscillatory spectrum may appear because of the term proportional
to $\cos\theta$ if $\theta$ depends on $k$.

\section{Summary and discussion}
\label{s5}
We studied the cosmological implications of quantum entanglement between two causally disconnected de Sitter spaces (BD1 and BD2) in the multiverse. As an initial state, we first considered an entangled state between BD1 and BD2, representing nucleation of a pair of universes, each either with one-particle states or in vacuum. We then derived the reduced density matrix of our universe (BD2) by tracing out the degrees of freedom of BD1 and examined the spectrum of vacuum fluctuations for BD2. We found that the quantum mechanical system becomes a mixed state by tracing out the inaccessible BD1, thus the spectrum turns out to consist of only diagonal elements.

In order to see the effect of entanglement, we considered a corresponding
non-entangled state and compared the spectrum. The correspondence was fixed 
by the condition that the probabilities for the vacuum and for the
one-particle states are the same if one focuses either on BD1 or BD2.
We then derived the reduced density matrix for BD2 similarly and examined 
the spectrum. If the state is non-entangled, the quantum mechanical system 
remains a pure state even after tracing out the inaccessible BD1. 
Consequently the off-diagonal elements remain.
This fact reflects on the spectrum. Namely a modulation may appear
in the spectrum due to quantum interference. Thus,
 the quantum interference distinguishes whether our universe is the
 entangled state or non-entangled state. We also mentioned that the 
modulation may contribute to the spectrum in different forms if we 
consider the bispectrum or higher order correlations, because the 
off-diagonal elements may contribute to the spectrum directly.
To demonstrate it by a simple example, 
we also considered an entangled state between the vacuum and the 
two-particle states.

We also mentioned that the degree of entanglement ${\cal C}$ may 
have a scale dependence. This is natural particularly from the
fact that the entangled state corresponds to truncation of an entangled 
state constructed by making use of the Bogoliubov transformation. 
Thus the modulation may have oscillatory behaviors. However, we 
should not be confused the oscillatory spectra as an effect of 
entanglement. 
Oscillatory spectra are always possible to obtain with an appropriate
 choice of the Bogoliubov coefficients between the Bunch-Davies vacuum
 and the entangled state~\cite{Albrecht:2014aga}. But such oscillations
are different from those due to quantum interference. In terms of the
coefficient ${\cal E}=|{\cal E}|e^{i\theta}$, 
the entangled state may have an oscillatory spectrum if $|{\cal E}|$ is 
oscillatory, while an oscillation may appear for the non-entangled state 
even if $|{\cal E}|$ has no oscillation but if its phase $\theta$ depends
on $k$.

What we have done is perhaps described in the simplest way that whether we choose a pure state or a mixed state as the initial state of our universe. This problem would be clarified if we get a more concrete picture of string landscape. In principle, it would be possible to distinguish between entangled and non-entangled states if we have means to observe
the $n$-point functions. Then this paper would offer a new set of vistas to be explored observationally in the future.

\section*{Acknowledgments}
I would like to thank Misao Sasaki and Jiro Soda for valuable discussions, 
suggestions and comments. 
I would also like to thank Jose Blanco-Pillado, Jaume Garriga, Takahiro Tanaka 
and Alex Vilenkin for careful reading of the draft, and for fruitful suggestions and comments. 
This work was supported by IKERBASQUE, the Basque Foundation for Science.

\appendix

\section{Entangled state of pairs of $n$-particles between BD1 and BD2}
\label{appA}

Here, we consider a squeezed state of pairs of $n$-particles between BD1 and BD2. As we show below, this state describes an entangled state between BD1 and BD2.

We focus on a single mode ${\bm k}$ of a massless minimally coupled scalar field in de Sitter space. The vacuum state for each BD1 and BD2 is defined as
\begin{eqnarray}
a_1\,|0\rangle_{\rm BD1}=0\,,\qquad
a_2\,|0\rangle_{\rm BD2}=0\,,
\end{eqnarray}
where $a_1$ and $a_2$ are annihilation operators for BD1 and BD2. Now let us consider a state $|\Psi\rangle$ defined by Bogoliubov transformations
\begin{eqnarray}
b_1=\alpha\,a_1+\beta\,a_2^\dagger\,,\qquad
b_2=\alpha\,a_2+\beta\,a_1^\dagger\,,
\end{eqnarray}
that is,
\begin{eqnarray}
b_1\,|\Psi\rangle=0\,,\qquad b_2\,|\Psi\rangle=0\,,
\end{eqnarray}
where $|\alpha|^2-|\beta|^2=1$. 
This state $|\Psi\rangle$ is then written by
\begin{eqnarray}
|\Psi\rangle=N\exp\left(-\frac{\beta}{\alpha}\,a_1^\dagger\,a_2^\dagger\right)
|0\rangle_{\rm BD1}|0\rangle_{\rm BD2}\,.
\label{squeezed1}
\end{eqnarray}
where $N$ is the normalization factor. This describes a squeezed state of pairs of $n$-particles between BD1 and BD2. If we expand the exponent in $|\Psi\rangle$, we have
\begin{eqnarray}
|\Psi\rangle&=&N\left(1-\frac{\beta}{\alpha}\,a_1^\dagger\,a_2^\dagger+\cdots\right)
|0\rangle_{\rm BD1}|0\rangle_{\rm BD2}\nonumber\\
&=&N\left(|0\rangle_{\rm BD1}|0\rangle_{\rm BD2}
-\frac{\beta}{\alpha}|1\rangle_{\rm BD1}|1\rangle_{\rm BD2}
+\frac{1}{2}\left(\frac{\beta}{\alpha}\right)^2|2\rangle_{\rm BD1}|2\rangle_{\rm BD2}+\cdots\right)\,.
\label{squeezed2}
\end{eqnarray}
This is an entangled state of the ${\cal H}_{\rm BD1}\otimes{\cal H}_{\rm BD2}$ space. If we truncate the above at the one particle state order, it is identical with the state in Eq.~(\ref{ini1}) and the degree of entanglement ${\cal C}$ corresponds to the Bogoliubov coefficients. Thus the degree of entanglement ${\cal C}$ may depend on the wavenumber.

\end{document}